\documentclass[prl,aps,showpacs,twocolumn,longbibliography]{revtex4-1}
\usepackage{bm}
\usepackage{amsmath}
\usepackage{amssymb}
\usepackage{graphicx}
\usepackage{slashed}
\usepackage{accents}
\usepackage{mathrsfs}
\usepackage{braket}
\usepackage{units}
\usepackage{upgreek}
\usepackage[caption=false]{subfig}

\usepackage[breaklinks=true]{hyperref}
\usepackage{breakurl} 

\DeclareSymbolFont{eulerscript}{U}{eur}{m}{n}
\DeclareSymbolFontAlphabet{\matheuler}{eulerscript}
\DeclareMathAlphabet{\boldgreek}{OML}{zplm}{b}{it}

\newcommand{\Q}{\mathcal{Q}}
\newcommand{\I}{i}
\newcommand{\nfrac}[2]{{#1}/{#2}}
\newcommand{\eps}{\epsilon}
\newcommand{\mc}[1]{\mathcal{#1}}
\newcommand{\spvec}[1]{\boldsymbol{#1}}
\newcommand{\lb}{\left(}
\newcommand{\rb}{\right)}
\newcommand{\rsb}{\right]}
\newcommand{\lsb}{\left[}
\newcommand{\rcb}{\right\}}
\newcommand{\lcb}{\left\{}
\newcommand{\abs}[1]{\left|#1\right|}
\DeclareMathOperator{\sign}{sign}
\DeclareMathOperator{\Ai}{Ai}
\newcommand{\Ftilde}{\mathfrak{F}}
\newcommand{\probsym}{W}
\newcommand{\airyx}{r}

\begin{document}
\title{High-Energy Recollision Processes of Laser-Generated Electron-Positron Pairs}
\author{Sebastian \surname{Meuren}}
\email{s.meuren@mpi-hd.mpg.de}
\author{Karen Z. \surname{Hatsagortsyan}}
\email{k.hatsagortsyan@mpi-hd.mpg.de}
\author{Christoph H. \surname{Keitel}}
\email{keitel@mpi-hd.mpg.de}
\author{Antonino \surname{Di Piazza}}
\email{dipiazza@mpi-hd.mpg.de}
\affiliation{Max-Planck-Institut f\"ur Kernphysik, Saupfercheckweg 1, D-69117 Heidelberg, Germany}
\date{\today}
\begin{abstract}
Two oppositely charged particles created within a microscopic space-time region can be separated, accelerated over a much larger distance, and brought to a recollision by a laser field. Consequently, new reactions become feasible, where the energy absorbed by the particles is efficiently released. By investigating the laser-dressed polarization operator, we identify a new contribution describing high-energy recollisions experienced by an electron-positron pair generated by pure light when a gamma photon impinges on an intense, linearly polarized laser pulse. The energy absorbed in the recollision process over the macroscopic laser wavelength corresponds to a large number of laser photons and can be exploited to prime high-energy reactions. Thus, the recollision contribution to the polarization operator differs qualitatively and quantitatively from the well-known one, describing the annihilation of an electron-positron pair within the microscopic formation region.
\end{abstract}

\maketitle

\begin{figure}[b]
\centering
\includegraphics{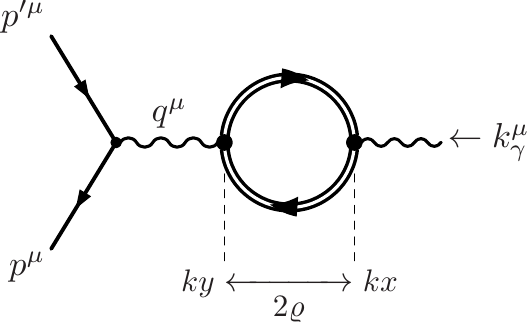}
\caption{\label{fig:recollisionfigure} 
Depending on the distance $2\varrho$ of the two polarization-operator vertices, this Feynman diagram describes either radiative corrections to the photon propagator or laser-induced recollision processes. The wavy lines denote photons, the double lines the laser-dressed electron (positron) propagators, and the straight solid lines indicate the particles produced in the secondary reaction. The meaning of the other symbols is explained in the text (time increases from right to left).}
\end{figure}

Recollision processes are responsible for a variety of phenomena, which have been investigated especially in the realm of atomic and molecular physics. After an atom (or a molecule) is ionized by a laser field, the electron is accelerated and possibly brought to a recollision with the parent ion. The energy that the electron absorbs between the ionization and the recollision can be released in different ways: as a high-energy photon after recombination [high-harmonic generation (HHG)] or by striking out another electron (nonsequential double ionization), see, e.g., \cite{salieres_study_1999,becker_above-threshold_2002,kohler_frontiers_2012,joachain_atoms_2012,di_piazza_extremely_2012}. The maximal energy absorbed by the recolliding ionized electron in a laser field with peak electric field strength $E_0$ and mean angular frequency $\omega$ is found to be about $3.17\,U_p$, where $U_p=\nfrac{e^2E_0^2}{(4m\omega^2)}$ is the ponderomotive potential, with $e<0$ and $m$ being the electron charge and mass, respectively \cite{kuchiev_atomic_1987,becker_higher-harmonic_1990,schafer_above_1993,corkum_plasma_1993,krausz_attosecond_2009}. Recently, applications of recollisions in nuclear physics have been discussed as well \cite{castaneda_cortes_nuclear_2013,lotstedt_nuclear_2014}.

Recollision processes also play an important role in high-energy physics as originating, for example, from an electron and a positron initially bound in a positronium atom, which may annihilate during a recollision and create other particles, analogously as in an ordinary collider experiment \cite{henrich_positronium_2004,hatsagortsyan_microscopic_2006,muller_muon_2008}. Moreover, the electron and the positron inducing the high-energy recollision process can also be created from vacuum in an ultrastrong laser field, e.g., in the presence of a laser field and a nucleus \cite{kuchiev_production_2007}. In both mentioned cases, classical considerations show that the available energy in the recollision, which happens after the particles propagated for approximately one laser wavelength, is of the order of $mc^2\xi^2=4U_p$, where $\xi=\nfrac{|e|E_0}{(m\omega c)}$. This explains why the ultrarelativistic regime $\xi\gg 1$ is of relevance in high-energy recollision physics. 

From a pictorial point of view one expects that in the realm of quantum field theory, recollisions are described by loop diagrams. The simplest Feynman diagram, which contains an electron-positron loop, is the polarization operator (PO); see Fig. \ref{fig:recollisionfigure}. Since the seminal work of Baier \textit{et al.} \cite{baier_interaction_1975} and Becker and Mitter \cite{becker_vacuum_1975}, the PO in a plane-wave field has been investigated in many publications (see, e.g., \cite{dittrich_probingquantum_2000,meuren_polarization_2013,
dinu_vacuum_2014,dinu_photon_2014,gies_laser_2014}). Surprisingly, no high-energy recollisions have been identified so far. To explain this, we note that for strong background fields ($\xi\gg 1$) the leading-order contribution to the PO in $\nfrac{1}{\xi}$ (quasistatic approximation) permits only the net exchange of a few laser photons \cite{di_piazza_refractive_2013}. It describes electron-positron pairs annihilating within the coherence length $\nfrac{\lambda}{\xi}$ of pair production, which is much smaller than the laser wavelength $\lambda=\nfrac{2\pi c}{\omega}$.

In the present Letter we consider high-energy recollisions experienced by an electron-positron pair, which is created by pure light in the collision of a gamma photon and an intense laser field (see Fig. \ref{fig:recollisionfigure}). It is shown for the first time that the PO contains subleading contributions in $\nfrac{1}{\xi}$, which allow for the efficient absorption of up to about $3.17\, \nfrac{\xi^3}{\chi}$ laser photons (if $\chi \gtrsim 1$), where $\chi=(2\nfrac{\hbar\omega_\gamma}{mc^2}) (\nfrac{E_0}{E_{\text{cr}}})$ is the quantum nonlinearity parameter, $\hbar\omega_\gamma$ the gamma photon energy and $E_{\text{cr}}= \nfrac{m^2c^3}{(|e|\hbar)} = \unitfrac[1.3\times 10^{16}]{V}{cm}$ the QED critical field  \cite{di_piazza_extremely_2012}. Contrary to the leading-order quasistatic approximation considered so far, the subleading contributions to the PO derived in this Letter describe recollision processes, as the creation and annihilation point of the electron-positron pair are separated on the scale set by the laser wavelength. Correspondingly, it is possible to absorb much more energy from the laser field (for $\chi\gtrsim 1$ the Heisenberg uncertainty relation is not violated, as even real $e^+e^-$ photoproduction is energetically allowed \cite{reiss_absorption_1962,nikishov_quantum_1964,ritus_1985,schutzhold_dynamically_2008,ehlotzky_fundamental_2009,dunne_catalysis_2009,bulanov_multiple_2010,ruffini_electronpositron_2010,heinzl_finite_2010,tuchin_non-linear_2010,orthaber_momentum_2011,ipp_streaking_2011,hebenstreit_pair_2011,krajewska_breit-wheeler_2012,titov_enhanced_2012,nousch_pair_2012,jiang_pair_2012,king_photon_2013,bulanov_electromagnetic_2013,jansen_strongly_2013,fedotov_pair_2013,titov_breit-wheeler_2013,lv_noncompeting_2013,lv_instantaneous_2014,kohlfurst_effective_2014,krajewska_breit-wheeler_2014,meuren_polarization-operator_2015}). Experimentally, the regime $\chi\gtrsim 1$, $\xi\gg1$ could be explored by colliding $\unit{GeV}$ photons (obtainable, e.g., via Compton backscattering) \cite{muramatsu_development_2014,leemans_multi-gev_2014,powers_quasi-monoenergetic_2014,wang_quasi-monoenergetic_2013,kim_enhancement_2013,phuoc_all-optical_2012,esarey_physics_2009,leemans_gev_2006} with strong optical laser pulses \cite{yanovsky_ultra_2008,ELI,CLF,XCELS}.

Exemplarily, we consider now recollision processes where the virtual photon decays into a lepton pair (see Fig. \ref{fig:recollisionfigure}). The generalization of the calculation to other secondary reactions is technically more involved but conceptually straightforward (from now on units with $\hbar=c=1$ and $\alpha = \nfrac{e^2}{4\pi}\approx \nfrac{1}{137}$ are used). 

To reduce the calculation to its essential part, we focus on the electron-positron loop and neglect the influence of the laser field on the final particles in Fig. \ref{fig:recollisionfigure}. After applying the usual Feynman rules \cite{peskin_introduction_2008,landau_quantum_1981}, the total recollision probability is given by
\begin{gather}
\label{eqn:recollision_decayratefinalvacuum}
\probsym(k_\gamma)
=
-  \int_{n_0}^\infty dn \, \frac{\eps_\mu \Pi^{\mu\nu}[\eps^\rho \Pi_{\rho\nu}]^*}{2kk_\gamma \alpha (2\pi)^2} \,\sigma_{\text{tot}}(q^2)
\end{gather}
(for an electron-positron pair in the final state also the scattering channel and the laser dressing must be taken into account). In Eq. (\ref{eqn:recollision_decayratefinalvacuum}) $k^\mu = (\omega, \spvec{k})$ denotes the average four-momentum of the laser photons, $k_\gamma^\mu = (\omega_\gamma, \spvec{k}_\gamma)$ and $\eps^\mu$ the momentum and polarization four-vector of the incoming gamma photon ($k_{\gamma}^2=0$), respectively, $\Pi^{\alpha\beta} = \Pi^{\alpha\beta}(k_\gamma,q)$ the nonsingular part of the PO (after renormalization of the vacuum part)  \cite{baier_interaction_1975,becker_vacuum_1975,dittrich_probingquantum_2000,meuren_polarization_2013,dinu_vacuum_2014,dinu_photon_2014,gies_laser_2014} and $q^\mu = k_\gamma^\mu + n k^\mu$ the four-momentum of the intermediate virtual photon ($\sqrt{q^2}= \sqrt{2n kk_\gamma}$ is the center-of-mass energy of the laser-induced electron-positron recollision). Furthermore, $\sigma_{\text{tot}}(q^2)$ represents the total cross section for the secondary process, e.g.,
\begin{gather}
\label{sigma_tot}
\sigma_{\text{tot}}(q^2) = \frac{4\pi \alpha^2}{3 q^2} \sqrt{1 - \nfrac{4m_\mu^2}{q^2}} \lb 1 + 2 \nfrac{m_\mu^2}{q^2} \rb
\end{gather}
for muon pair production \cite{peskin_introduction_2008,landau_quantum_1981} and $n_0$ a possible kinematic threshold [e.g. $n_0 = 2\nfrac{m_\mu^2}{(kk_\gamma)}$ for muon pair production, where $m_\mu$ is the muon mass].

Recollisions are only possible if the laser is linearly polarized. Therefore, we use $A^\mu(\phi) = a^\mu \psi(\phi)$ for the four-potential ($\xi = \nfrac{\abs{e}\sqrt{-a^2}}{m}$, $\chi=\xi\, \nfrac{kk_\gamma}{m^2}$) and $\Ftilde^{\mu\nu}(\phi,\phi_0) = \int^{\phi}_{\phi_0} d\phi'\, F^{\mu\nu}(\phi')=f^{\mu\nu} [\psi(\phi)-\psi(\phi_0)]$ for the integrated field-strength tensor, where $f^{\mu\nu} = k^\mu a^\nu - k^\nu a^\mu$. In the numerical calculations we have used the pulse $\psi'(\phi) = \sin^2[\nfrac{\phi}{(2N)}] \, \sin(\phi)$ with $N=5$ cycles (the prime denotes the derivative with respect to the argument).

\begin{figure}
\centering
\includegraphics{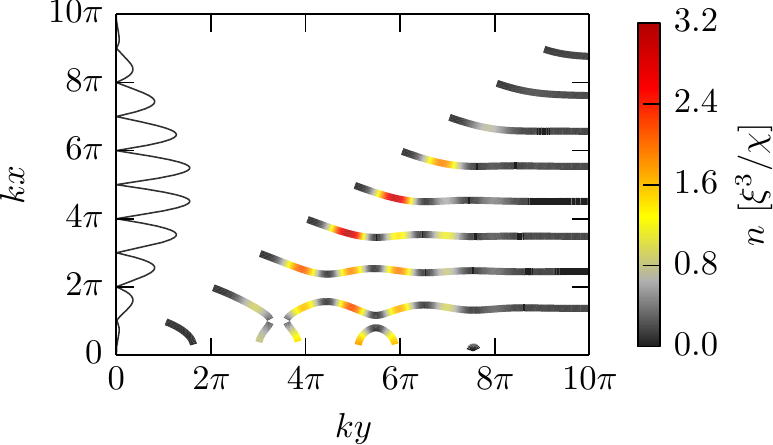}
\caption{\label{fig:recollision_qsps}(color online) Combinations of creation ($kx$) and annihilation ($ky$) laser phases for which a recollision is possible [see Eq.~(\ref{eqn:recollision_classicalrecollisioncondition})]. The color depicts the quantity $n$ proportional to the energy absorbed (classically) by the electron-positron pair in the laser field [see Eq.~(\ref{eqn:recollision_classicalrecollisionenergy})]. The solid black line illustrates the absolute value of the laser-envelope function $\psi'(kx)$ in arbitrary units.}
\end{figure}

For linear polarization the field-dependent part of the PO reads (the notation agrees with \cite{meuren_polarization-operator_2015})
\begin{multline}
\label{eqn:recollision_linpol_realincoming}
\Pi^{\mu\nu}(k_{\gamma},q)
-
\Pi^{\mu\nu}_{\Ftilde=0}(k_{\gamma},q)
= 
- \int_{-\infty}^{+\infty} dky \, \int_0^{\infty} \frac{d\varrho}{\varrho}
\\\times\,
\frac{\alpha}{2\pi} \, \big[
P_{11} \Lambda_1^\mu \Lambda_1^\nu
+
P_{22} \Lambda_2^\mu \Lambda_2^\nu
+
P_{Q} \Q_1^\mu \Q_2^\nu
\big],
\end{multline}
where $\varrho = \nfrac{(ky-kx)}{2}$ and $kx$ ($ky$) denotes the laser phase when the pair is created (annihilated) (see Fig. \ref{fig:recollisionfigure}). In the considered case of an incoming photon, counterpropagating with respect to the laser field, the four-vectors $\Lambda^\mu_1$ and $\Lambda^\mu_2$ can be chosen as $\eps_{\parallel}^\mu$ and $\eps_{\perp}^\mu$, respectively (the indexes $\parallel$ and $\perp$ refer to the polarization direction of the laser), whereas the last term on the right-hand side of Eq. (\ref{eqn:recollision_linpol_realincoming}) does not contribute.  Thus, $\abs{\int P_{\perp,\parallel}}^2=\abs{\int_{-\infty}^{+\infty} dky \, \int_0^{\infty} d\varrho\,\varrho^{-1} \, P_{\perp,\parallel}}^2$ determines the recollision probability for the corresponding photon polarization [see Eq. (\ref{eqn:recollision_decayratefinalvacuum})]. For the subsequent analysis it is sufficient to note that the coefficients $P_{11} = P_{\parallel}$ and $P_{22} = P_{\perp}$ contain the oscillatory phase factor $\exp{[\I\varphi(\varrho,ky)]}$ (see \cite{meuren_polarization-operator_2015} for details), where 
\begin{gather}
\label{eqn:recollisions_fielddependentphase}
\varphi(\varrho,ky)
=
n ky
- 
4 \, (\nfrac{\xi}{\chi}) \, [\varrho + \xi^2 D(\varrho,ky)],
\end{gather}
with $D(\varrho,ky)  = \varrho (J-I^2)$ and
\begin{gather}
\label{eqn:recollision_IJdefinition}
I
=
\int_{0}^{1} dl\, \psi(ky - 2\varrho l),
\quad
J
=
\int_{0}^{1} dl\, \psi^2(ky - 2\varrho l).
\end{gather}
We first investigate the integral in $\varrho$ for a fixed value of $ky$. For $\xi \gg 1$ and at fixed $\chi$ the phase factor $\exp{[-4\I  \, (\nfrac{\xi^3}{\chi}) \, D(\varrho,ky)]}$ is highly oscillating and we can apply a stationary-phase analysis. The stationary points $\varrho_k$ are determined by the condition $D'(\varrho_k,ky)=[\psi(ky - 2 \varrho_k) - I(\varrho_k,ky)]^2=0$, which implies
\begin{gather}
\label{eqn:recollision_classicalrecollisioncondition}
(ky-kx) \psi(kx) = \int_{kx}^{ky} d\phi \, \psi(\phi)
\end{gather}
(for $D$ the prime denotes the partial derivative with respect to $\varrho$). Equation (\ref{eqn:recollision_classicalrecollisioncondition}) links the creation ($kx$) and the annihilation ($ky$) phases (see Fig.~\ref{fig:recollision_qsps}) and it has a straightforward classical interpretation. In fact, the electron ($P_{+}^\mu$) and positron ($P_{-}^\mu$) four-momentum $P_{\pm}^\mu(\phi)=[\mathcal{E}_{\pm}(\phi),\bm{P}_{\pm}(\phi)]$ inside a plane-wave field is classically given by \cite{di_piazza_extremely_2012,sarachik_classical_1970}
\begin{gather}
\label{eqn:recollision_momentumfourvectorsolution}
P_{\pm}^\mu(\phi)
=
P_{\pm,0}^\mu \pm \frac{e \Ftilde^{\mu}_{\phantom{\mu}\nu}(\phi,\phi_0)  P^{\nu}_{\pm,0}}{kP_{\pm,0}} 
+
\frac{e^2 \Ftilde^{2\mu}_{\phantom{2\mu}\nu}(\phi,\phi_0) P^{\nu}_{\pm,0}}{2(kP_{\pm,0})^2},
\end{gather}
where $P_{\pm,0}^\mu = P_{\pm}^\mu(\phi_0)$. To determine the recollision point from the classical trajectory, we assume that both particles are created at the same space-time point $x^\mu$ (laser phase $\phi_0=kx$). The more asymmetrically the initial photon momentum is distributed between the two created particles, the smaller becomes the recollision probability (as the impact parameter increases). Therefore, we also assume that both particles have the same initial momentum $\spvec{P}_{\pm,0} = \nfrac{\spvec{k}_\gamma}{2}$ and we find that Eq.~(\ref{eqn:recollision_classicalrecollisioncondition}) exactly corresponds to the condition that the (classical) coordinates of the electron and the positron coincide again at a later phase $\phi=ky$. Since one needs $\omega_\gamma \gg m$ to reach $\chi \gtrsim 1$, the particles are created with ultrarelativistic energies.
 
\begin{figure}
\centering
\includegraphics{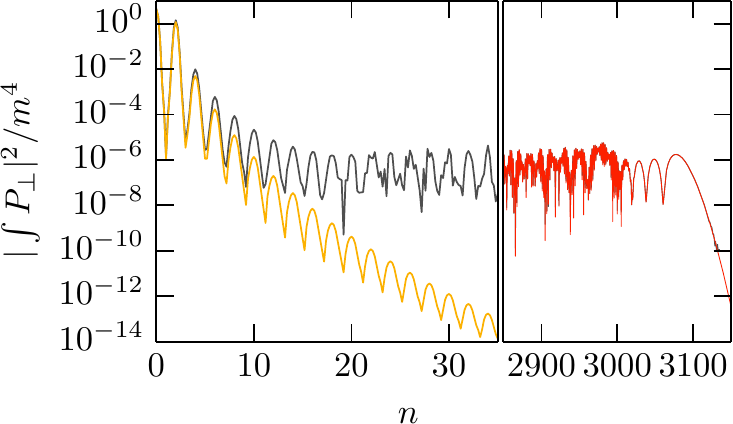}
\caption{\label{fig:recollision_comparison}(color online) Left side: Comparison of the quasistatic contribution (lower yellow curve) with the full numerical calculation (upper gray curve). Right side: Plateau-region, analytical (red curve) and numerical calculation coincide [$\chi=1$, $\xi=10$, $N=5$, see Eq.~(\ref{eqn:recollision_linpol_realincoming})].}
\end{figure}

The stationary-phase equation $D'(\varrho_k,ky)=0$ always admits the solution $\varrho_0=0$, independently of the shape of the background field. The contribution of this stationary point is formed for values of $\varrho$ in the region $0\le\varrho\lesssim \nfrac{1}{\xi}$, where the phase $4\,(\nfrac{\xi^3}{\chi})\,D(\varrho,ky)$ is less than or of the order of unity. Thus, this contribution describes the immediate annihilation of the created electron-positron pair within a distance of the order of $\nfrac{\lambda}{\xi}$ inside an (effectively) constant-crossed field (quasistatic limit). The compensation of the large parameter $\nfrac{\xi^3}{\chi}$ in the phase occurring at $\rho\lesssim \nfrac{1}{\xi}$ explains why the stationary point $\rho_0$ provides the leading contribution to the PO and, at the same time, why it allows for a net exchange of only a few laser photons \cite{di_piazza_refractive_2013}. On the other hand, laser-induced recollision processes are described by the contributions to the integral in $\varrho$ close to the nonvanishing stationary points $\varrho_k$, $k=1,2,\ldots$, with $\varrho_k\gtrsim \pi\gg \nfrac{1}{\xi}$ (see Fig. \ref{fig:recollision_qsps}). As we will see below, these contributions are formed in the regions $|\varrho-\varrho_k|\lesssim \nfrac{1}{\xi}$, where the phase $4\,(\nfrac{\xi^3}{\chi})\,D(\varrho,ky)$ remains of the order of $\nfrac{\xi^3}{\chi}$. Thus, although such contributions are suppressed with respect to that from $\varrho_0$, they are essential to understand the high-energy plateau region of the photon-absorption spectrum [see Fig.~\ref{fig:recollision_comparison} (left side)].

In order to determine the contribution from the recollision processes, we expand the function $D(\varrho,ky)$ around $\varrho_k$ up to the third order [$D''(\varrho_k,ky)=0$], with $D'''(\varrho_k\neq0,ky)  = 8 \big[ \psi'(ky - 2\varrho_k) \big]^2$. Since the third-order term of the expansion scales as $\nfrac{(\varrho - \varrho_k)^3\xi^3}{\chi}$, the contribution is formed in the region $|\varrho-\varrho_k|\lesssim\nfrac{1}{\xi}$ and also the linear term in $\xi$ in $\varphi(\varrho,ky)$ must be taken into account. All higher-order terms can be neglected. On the other hand, the preexponent functions in $P_{\perp,\parallel}$ vanish at $\varrho_k$ and it is necessary to expand them up to linear terms in $\varrho-\varrho_k$. Now, we apply the change of variable $\varrho - \varrho_k=[r\nfrac{\chi}{(4\xi)}]t$, where $r = [\nfrac{2}{\chi(\varrho_k,ky)}]^{\nicefrac{2}{3}}$, with $\chi(\varrho,ky) = \chi \abs{\psi'(ky - 2\varrho)}$ being the quantum-nonlinearity parameter at the pair-production vertex. Then, the phase $\varphi(\varrho,ky)$ can be approximated by $\varphi(\varrho,ky) \approx \varphi(\varrho_k,ky) - (t \airyx + \nfrac{t^3}{3})$. %
After extending the integration boundaries in the new variable $t$ to $\pm \infty$, the contribution to the integral from the region around the stationary point $\varrho_k \neq 0$ reads
\begin{multline}
\label{eqn:recollision_varrhomasterint}
\int_0^\infty d\varrho \, g(\varrho) e^{\I \varphi(\varrho,ky)} 
\approx 
e^{\I \varphi(\varrho_k,ky)} \, \nfrac{\pi\chi}{(2\xi)} 
\\\times \, \big[ g(\varrho_k)\, \airyx \Ai(\airyx) 
+ \I g'(\varrho_k) \Ai'(\airyx)\, \nfrac{\airyx^2\chi}{(4\xi)}  \big],
\end{multline}
where $g(\varrho)$ is an arbitrary, slowly varying function and $\Ai$ is the Airy function \cite{olver_nist_2010}. As expected, at $\chi\ll 1$ the above contribution features an exponential suppression $\sim \exp{\lcb-\nfrac{4}{[3\chi(\varrho_k,ky)]}\rcb}$, i.e. as the electron-positron pair-production amplitude inside a (locally) constant-crossed field \cite{di_piazza_extremely_2012,ritus_1985}.

By applying Eq.~(\ref{eqn:recollision_varrhomasterint}) to Eq.~(\ref{eqn:recollision_linpol_realincoming}) we obtain
\begin{subequations}
\label{eqn:recollision_varrhointegrals}
\begin{multline}
\int_0^\infty \frac{d\varrho}{\varrho} \,  P_{\perp}
\approx \I m^2  \, e^{\I \varphi(\varrho_k,ky)} \, \frac{\pi\chi^2}{2} \,
\Big\{
\psi'(ky-2\varrho_k) 
\\\times M(\varrho_k,ky) 
\mc{W}_0 [x_1(\varrho_k,ky)]  \, \nfrac{\airyx^2 \Ai'(\airyx)}{(4\varrho_k)}
\\
- \mc{W}_2 [x_1(\varrho_k,ky)] \, \nfrac{\airyx \Ai(\airyx)}{(\varrho^2_k \xi^2)} \Big\}
\end{multline}
and
\begin{multline}
\int_0^\infty \frac{d\varrho}{\varrho} \,  P_{\parallel}
\approx 
- \I m^2  \, e^{\I \varphi(\varrho_k,ky)} \, \psi'(ky-2\varrho_k)  M(\varrho_k,ky)
\\\times 
\mc{W}_1 [x_1(\varrho_k,ky)]\, \airyx^2 \Ai'(\airyx) \frac{\pi\chi^2}{2\varrho_k} + \int_0^\infty \frac{d\varrho}{\varrho} \,  P_{\perp},
\end{multline}
\end{subequations}
where $x_1(\varrho,ky) = (\nfrac{\xi}{\chi}) \, [\varrho + \xi^2 D(\varrho,ky)]$,  $M(\varrho,ky) = \psi(ky) - \psi(ky-2\varrho)$ and the functions $\mc{W}_i$ are defined in \cite{meuren_polarization-operator_2015}.

\begin{figure}
\centering
\includegraphics{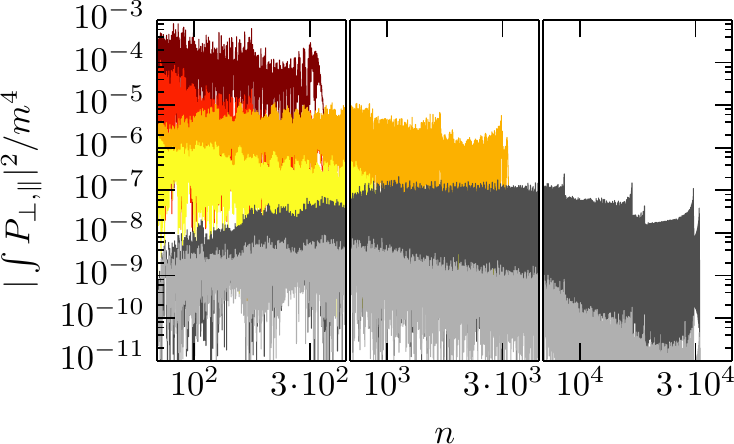}
\caption{\label{fig:recollision_scaling}(color online) Numerically calculated photon absorption spectra for $\chi=1$, $N=5$ and $\xi=10^{\nfrac{2}{3}} \approx 4.6$ (upper two), $\xi=10$ (middle two), $\xi=10^{\nfrac{4}{3}} \approx 21.5$ (lower two). Of each pair the upper (lower) spectrum corresponds to perpendicular (parallel) polarization.}
\end{figure}

Now, we proceed to determine the stationary points of the integral in $ky$. To this end we have to solve the equation $ \varphi'(ky) = 0$, where $\varphi(ky) = \varphi[\varrho_k(ky),ky]$  [see Eq.~(\ref{eqn:recollision_varrhomasterint})]. The stationary-phase condition reads
\begin{gather}
\label{eqn:recollision_statphaseforky}
n  
= 
\frac{\xi^3}{\chi} \bigg[ 2M^2(\varrho_k,ky) + \frac{2}{\xi^2} - \frac{1}{\xi^2} \frac{M(\varrho_k,ky)}{\varrho_k \psi'(ky - 2\varrho_k)} \bigg],
\end{gather}
where $\varrho_k = \varrho_k(ky)$ [note that for $\psi'(ky - 2\varrho) \to 0$ the pair-production probability is exponentially suppressed]. The leading-order contribution with $n \approx 2M^2(\varrho_k,ky) \nfrac{\xi^3}{\chi}$ corresponds to the four-momentum 
\begin{gather}
\label{eqn:recollision_classicalrecollisionenergy}
k'^\mu  = n k^\mu = 2 (\nfrac{\xi^3}{\chi}) [\psi(ky)-\psi(kx)]^2 \, k^\mu
\end{gather}
that the electron-positron pair has classically absorbed from the laser field [see Eq.~(\ref{eqn:recollision_momentumfourvectorsolution}) and Fig.~\ref{fig:recollision_qsps}]. For a monochromatic field pairs created after the peak ($\phi\approx\nfrac{\pi}{10}$) have the highest recollision energy and we obtain the cutoff $n_c = 3.17\, \nfrac{\xi^3}{\chi}$, which corresponds to the result $3.17\, U_p$ obtained in atomic HHG ($U_p = \nfrac{m\xi^2}{4}$) \cite{kuchiev_atomic_1987,becker_higher-harmonic_1990,schafer_above_1993,corkum_plasma_1993,di_piazza_extremely_2012}. The $\nfrac{\xi^3}{\chi}$ scaling of the cutoff is confirmed by a full numerical calculation in Fig. \ref{fig:recollision_scaling}. 

Now, the approximate contribution of the stationary point $ky_s$, solution of Eq.~(\ref{eqn:recollision_statphaseforky}), is given by
\begin{gather}
\label{eqn:recollision_kystatpointcontribution}
\int_{-\infty}^{+\infty} dky \, h(ky) \, e^{\I \varphi(ky)}
\approx 
h_s \, e^{\I (\sigma\nfrac{\pi}{4}+\varphi_s)} \sqrt{\frac{2\pi}{\abs{\varphi''_s}}},
\end{gather}
where $h(ky)$ is an arbitrary, slowly-varying function, $h_s=h(ky_s)$, $\varphi_s=\varphi(ky_s)$, $\varphi''_s=\varphi''(ky_s)$, $\sigma=\sign(\varphi''_s)$, and
\begin{gather}
\label{eqn:recollision_phasesecondderivativeky}
\varphi''(ky) 
=
- 4 \frac{\xi^3}{\chi} \lsb M(\varrho_k,ky) \psi'(ky) - \frac{M^2(\varrho,ky)}{2\varrho_k} + \frac{\varrho''_k}{\xi^2} \rsb,
\end{gather}
where $\varrho_k'' = \varrho_k''(ky)$. If two stationary points coalesce, the Airy uniform approximation \cite{chester_extension_1957} must be used instead of Eq.~(\ref{eqn:recollision_kystatpointcontribution}). In Fig. \ref{fig:recollision_comparison} the recollision contribution was calculated analytically using the above approximations and compared with a full numerical calculation (right side). For large photon numbers both results agree already for $\xi=10$.

Combining Eqs. (\ref{eqn:recollision_varrhointegrals}) and (\ref{eqn:recollision_kystatpointcontribution}), we conclude that the recollision contribution to $\abs{\int P_{\perp}}^2$ scales as $\xi^{-6}\chi^{10/3}$ at $\chi\gtrsim 1$ [$\mc{W}_i(x) \sim x^{-\nfrac{1}{2}}$ \cite{meuren_polarization-operator_2015}, the $\xi^{-6}$-scaling is confirmed numerically in Fig. \ref{fig:recollision_scaling}], while the quasistatic contribution is independent of $\xi$ \cite{meuren_polarization_2013}.

Using Eq. (\ref{eqn:recollision_decayratefinalvacuum}) we can estimate the $\xi$ scaling of the ratio $R$ between the recollision probability $\probsym(k_\gamma)$ and the pair-production probability $\probsym_{e^+e^-}(k_\gamma)$. Since $n\sim \xi^3$, $\eps_\mu \Pi^{\mu\nu}[\eps^\rho \Pi_{\rho\nu}]^*\sim m^4\xi^{-6}$, and $\probsym_{e^+e^-}(k_\gamma) \sim \xi$ \cite{meuren_polarization-operator_2015}, we obtain $R \sim \xi^{-3} m^2 \sigma_{\text{tot}}$, where $\sigma_{\text{tot}}\sim \sigma_{\text{tot}}(q^2=m^2\xi^2)$. An intuitive explanation of this scaling is based on the wave packet spreading of the electron-positron pair between the production and the recollision as in \cite{hatsagortsyan_microscopic_2006,kuchiev_production_2007}. In fact, in the frame where the recollision is head-on and along the polarization direction of the laser, one obtains $R\sim \nfrac{\sigma_{\text{tot}}}{\mc{A}}$ \cite{peskin_introduction_2008}, where $\mc{A} = \Delta L_{b} \Delta L_{k}$, with $\Delta L_b\sim \nfrac{\xi}{m}$ and $\Delta L_k\sim\nfrac{\xi^2}{m}$ being the spread of the particles along the direction of the magnetic field and along the propagation direction, respectively [see also Eq. (\ref{eqn:recollision_momentumfourvectorsolution})]. Note that the initial conditions assumed in \cite{kuchiev_production_2007} for the classical propagation do not correspond to the most probable ones employed here, which explains the different scaling of the quantity $R$ there.

As the tree-level production of $\mu^+\mu^-$ pairs is exponentially suppressed [$\exp(-\nfrac{8}{3\chi_{\mu}})$, with $\chi_{\mu}\sim 10^{-7} \chi$], their observation would unambiguously prove the existence of recollision processes. For a pulse with $5$ cycles, $\chi=1$ and $\xi = 200 \approx \nfrac{m_\mu}{m}$ (kinematic threshold) we obtain to leading order a probability of $2 \times 10^{-20}$ per incoming gamma photon. However, for $\xi \gtrsim 200$ the emission of additional photons within the electron-positron loop should be taken into account. In fact, $N_\gamma \approx \pi\alpha\xi$ photons are emitted on average by each particle in the loop (integrating the total emission probability yields $N_\gamma=4.5$) \cite{di_piazza_extremely_2012}. Taking the corresponding exponential decay of the electron (positron) wave function into account \cite{meuren_quantum_2011,di_piazza_extremely_2012}, the probability for $\mu^+\mu^-$ pair production without the emission of additional photons is $\sim 10^{-24}$. We stress that this is a lower bound for the exact probability, as the emission of additional soft photons does not prevent the recollision.

Finally, we consider the case of an $e^+e^-$ pair in the final state. Because of the $\xi^{-6}$ scaling of the plateau (see Fig. \ref{fig:recollision_scaling}) much higher recollision probabilities are now obtained in the regime $\chi=1$, $\xi=10$. From Eq. (\ref{eqn:recollision_decayratefinalvacuum}) we expect $\sim 10^{-13}$ recollision events per incoming gamma photon for a $5$-cycle pulse. Furthermore, in this case $N_\gamma<1$ and still the absorption of $\sim 10^3$ laser photons is possible. Therefore, in comparison with tree-level $e^+e^-$ photoproduction, recollision-produced $e^+e^-$ pairs cover a much wider phase-space region, such that these channels are, in principle, distinguishable (after the recollision also the momentum component $p_B$ of each particle along the direction of the laser magnetic field spans up to $m\xi$, whereas for the tree-level channel momenta $p_B > m$ are exponentially suppressed \cite{ritus_1985}).

\begin{acknowledgments}
A.D.P. would like to thank Alexander I. Milstein and S.M. Andreas Fischer, Andreas Kaldun, Ben King, Anton W\"ollert and Enderalp Yakaboylu for fruitful discussions. S.M. is also grateful to the Studienstiftung des deutschen Volkes for financial support. All plots have been created with \textsc{Matplotlib} \cite{hunter_matplotlib_2007} and the \textsc{GSL} \cite{GSL} has been used for numerical calculations.
\end{acknowledgments}

\end{document}